\documentclass[12pt]{article}
\pdfoutput=1
\usepackage{graphicx,amssymb,amsfonts,amsthm,amsmath,amsbsy,amscd,amstext, color}
\usepackage{subcaption}
\usepackage{jheppub}

\newcommand{\e}{\epsilon}
\newcommand{\del}{\partial}

\def\cos{\operatorname{cos}}
\def\sin{\operatorname{sin}}

\def\upr{u_\perp}
\def\upl{u_\parallel}

\title{Phase Transitions and Conductivities of Floquet Fluids}

\preprint{\today}

\author[a]{Andrew Baumgartner} 
\author[a]{ and Michael Spillane}
\affiliation[a]{Department of Physics, University of Washington, Seattle, WA, 98195-1560, USA}
\emailAdd{baum4157@uw.edu}
\emailAdd{spillan3@uw.edu}

\abstract{We investigate the phase structure and conductivity of a relativistic fluid in a circulating electric field with a transverse magnetic field.  This system exhibits behavior similar to other driven systems such as strongly coupled driven CFTs \cite{Rangamani2015} or a simple anharmonic oscillator. We identify distinct regions of fluid behavior as a function of driving frequency, and argue that a ``phase" transition will occur. Such a transition could be measurable in graphene, and may be characterized by sudden discontinuous increase in the Hall conductivity. The presence of the discontinuity depends on how the boundary is approached as the frequency or amplitude is dialed.  In the region where two solution exists the measured conductivity will depend on how the system is prepared.
}

\begin{document}
\maketitle

\section{Introduction}
The dynamics of periodically driven systems is a rich and interesting problem. Weakly coupled systems subject to periodic driving develop topological phases due to extra periodicity constraints imposed on the Hamiltonian. Such ``quasi-energies" can change the topology of the band structure, leading to topologically non-trivial phases \citep{Cayssol:2013, Carpentier:2015, Roy:2017, Rudner:2015}. An interesting example of this is the Floquet-Weyl semi-metal \citep{WSM:1,WSM:2,WSM:3}, where a free Dirac cone in a rotating electric field, splits into two Weyl cones. The separation distance is proportional to the field strength over the driving frequency, and so such a transition can be studied as a function of driving frequency. Other interesting examples include Floquet time crystals \citep{TimeCrystals}, Floquet symmetry protected topological phases \citep{FSPT} and symmetry protected Floquet quasi-particles \cite{Lukasz:2,Rudner:2013}. Common in all of these systems is the emergence of new phases as a function of frequency; the system behaves drastically differently depending on how fast you drive it. 

Things remain interesting at strong coupling, although one is required to abandon the standard band topology picture. Nevertheless it is still possible to identify distinct phases as a function of driving frequency. For example, the authors of \citep{Rangamani2015} studied a holographic system driven by a periodic scalar operator. They found that for weak amplitudes and fast driving the dynamics is dominated by dissipation, while the other extreme exhibits unbounded amplification (see also \cite{Biasi:2017kkn, Auzzi:2013pca} for similar set ups) . Similar results have been found for holographic superconductors driven by an oscillating electric field in the $x$-direction \cite{Li:2013fhw} and for probe brane holography in 3+1 dimensions \cite{Hashimoto2017, Kinoshita:2017uch}.  

Crucial to these examples is the full non-linearities of Einstein's equations. This motivates the question:  Does one still find a robust phase diagram without including the full non-linearities of Einstein's equations? Recent work on periodically driven, non-relativistic fluids \citep{FloquetHydro} has taken a step towards answering this question. The authors of \citep{FloquetHydro} have found the formation of a boundary-layer fluid as a function of driving frequency. Such a boundary layer phase is not present in the limit of low driving, offering an example of distinct frequency dependent phases for fluids. 

In this work, we consider the dynamics of a relativistic fluid in 2+1 dimensions in a magnetic and circularly polarized electric field. Unlike \citep{FloquetHydro}, we do not place our fluid in a box, but nonetheless find distinct phases characterized by the value driving frequency. Namely, we find two distinct regions, characterized by monotonically increasing and decreasing conductivity separated by a region of instability.  We hope that such a transition can be experimentally realized in the Dirac fluid phase of graphene \citep{GrapheneHydro} using a generalization of the techniques proposed in \citep{Graphene}. 

The paper is structured as follows. In section 2 we provide a motivating example of a relativistic particle in a magnetic field subject to periodic driving. In section 3 we present our hydrodynamic set up, as well as an exact solution for the fluid equations and analytical results for conductivities in various limits. In section 4 we subject our system to an external DC probe and calculate the conductivities. Here we will present our numerical techniques and results. We conclude in section 5. Various technical details are collected in the appendix. 

\section{Motivating example} \label{sec:example}
We are interested in relativistic fluids which, in the low density limit, should bear a resemblance to a dilute non-interacting gas.  If we can find a periodic solution for a single particle for a given field configuration we expect that a similar solution will exist for a fluid. We consider a charged particle confined to a plane in an oscillating electric field, similar to circularly polarized light \footnote{This field configuration is not a solution to Maxwell's equations, however we assume the forces generated by the components of the magnetic field tangent plane are canceled by the forces that keep the particle in the plane.}  

\begin{align}
F=E \cos(\omega t) dt\wedge dx + E \sin(\omega t) dt\wedge dy+B dx\wedge dy \, .
\end{align}

The Lorentz equation can be written covariantly as

\begin{align}
\frac{d u_\mu}{d\tau} = \frac{q}{m}F_{\mu \nu} u^\nu \, .
\end{align}
We are interested in particles with closed trajectories.  Such particles have constant energy and so we can make the replacement $\tau \rightarrow t/\gamma$.  With this simplification, and a simple ansatz, we can find $\upr$ is given by the solution to

\begin{align} \label{particle}
u_\mu = \{-\sqrt{1+\upr^2},-\upr\sin(\omega t),\upr\cos(\omega t)\} \, ,\\
B q \upr=\sqrt{1+\upr^2}(E q +m\upr\omega)\, .
\end{align}
In the $\omega \to 0$ limit, assuming $E<B$ and solving for $\upr$, we find 
\begin{equation}
\upr=\frac{qE}{\sqrt{B^2-E^2}}
\end{equation}
while for $B<E$, a real solution for $\upr$ does not exist. Similarly, for $\omega \to \infty$, we find that $\upr=0$. Intuitively this can be thought of as the massive particle never being able to ``catch up" to the driving, since it does not have sufficient amount of time to react to the external force. We expect to see qualitatively similar behavior in the relativistic fluid setting.

\section{Floquet Fluid Solution}

Let us extend the dilute gas example given above to a perfect fluid.  The stress tensor for a perfect fluid is written as 

\begin{align}
T^{\mu\nu} = (\epsilon + p) u^\mu u^\nu +p  \eta^{\mu\nu} \, 
\end{align}
where $\epsilon$ is the energy density and $p$ is the pressure.  The equations of motion in the lab frame are then given by

\begin{align}
\label{T_cons}\nabla_\mu T^{\mu\nu} &= F^{\mu\nu}J_\mu \, +\frac{1}{\tau}(\delta^\nu_\mu+n^\nu n_\mu)T^{\mu\lambda}n_\lambda, \\
\label{J_cons}\nabla_\mu J^\mu &= 0 \, ,
\end{align}
where $n^\mu=\{-1,0,0\}$ is a unit vector in the $t$ direction.  The impurities are static in the lab frame and therefore pick a preferred time direction $n^\mu$.  \footnote{ Some recent papers have used the term $\frac{1}{\tau}(\delta^\nu_\mu+u_\mu u^\nu)T^{\mu\lambda}u_\lambda$, as opposed to $\frac{1}{\tau}(\delta^\nu_\mu+n^\nu n_\mu)T^{\mu\lambda}n_\lambda$ as a covariant way to write the momentum relaxation term. This is identically zero and therefore does not have any effect on the momentum.} One should not expect a covariant expression for momentum relaxation because the impurities explicitly break Lorentz invariance. 

Note that for 2+1 dimensional fluids with a magnetic field there is no longer a delta function in the DC conductivity. The magnetic field breaks the boost invariance which was responsible for the delta function. It can be attributed to the fact that a boost of the static, zero velocity solution creates a current with no electric field.  This would only be possible if the conductivity was infinite. We will consider the case where there is no impurity scattering.

We consider the case where the current is $J^\mu = \rho u^\mu -\sigma_Q F^{\mu\nu}u_\nu$, the charge density is $\rho$ and $F_{\mu\nu}$ is field strength for the external electric and magnetic field and is given by

\begin{align}
F=E \cos(\omega t) dt\wedge dx + E \sin(\omega t) dt\wedge dy+B dx\wedge dy.
\end{align}
Implicit in our choice of $J^\mu$ is the assumption that $E,B \propto \mathcal{O}(\partial)$, so that we may consistently truncate the constitutive relations to first order in $F_{\mu \nu}$ \citep{Baumgartner:2017kme}. 

In the ideal fluid case with $\sigma_Q =0$ and $\tau\rightarrow \infty$ there is a simple stationary solution where $E\cdot J$ =0. The solution is given as

\begin{align}
u_\mu = \{-\sqrt{1+\upr^2},-\upr\sin(\omega t),\upr\cos(\omega t)\} \, ,\\
\label{sQ0} B \rho \upr=\sqrt{1+\upr^2}(E \rho +(\epsilon+p)\upr\omega)\, ,
\end{align}
where $\epsilon$ and $p$ are constant. This is exactly analogous to the dilute gas solution presented in Section \ref{sec:example} in the ``quasi-particle limit", i.e. $\epsilon+p \to m$ and $\rho \to q$. 

The equation for $\upr$ can be solved, but the closed form is messy and not particularly illuminating except to notice that there is no divergence as $\omega$ approaches the cyclotron frequency. In fact, looking for solutions where $\upr \rightarrow \infty$ we see there are no solutions with $\omega \neq 0$ for finite $\rho$, $E$ and $B$.

To proceed, we first assume the existence of a heat sink into which our fluid can dump excess energy. This allows us to treat the energy as constant even though the electric field is continually pumping energy into the system. With this simplification, we can find the fluid velocity in terms of two constants.  Using the ansatz

\begin{align}
u_x = -\upr \sin(\omega t)+\upl \cos(\omega t) \, \\
u_y = \upr \cos(\omega t)+\upl \sin(\omega t) \, 
\end{align}
we find that $\upr$ and $\upl$ must satisfy 
\begin{align}
\label{fluid1} u_t \left( -\frac{1}{\tau}\upr  + \upl \omega - \sigma_Q \frac{B E}{\epsilon+p} \right)-\upr \sigma_Q \frac{B^2}{\epsilon+p} +\upl \frac{\rho B}{\epsilon+p}=0 \, ,\\
\label{fluid2}   u_t \left( \frac{1}{\tau} \upl +\frac{\rho E}{\epsilon+p}+ \upr \omega \right) -\upl \sigma_Q  \frac{\left(E^2-B^2\right)}{\e+p} + \upr \frac{\rho B}{\epsilon+p}= 0 \, .
\end{align}
We can rewrite the above in terms of the dimensionless quantity $\tilde{\omega}\equiv\omega (\epsilon+p) / \rho B =\omega /\omega_B$ to obtain

\begin{align}
\label{dim_fluid1} u_t \left( -\frac{1}{\tau \omega_B} \upr  + \upl \tilde{\omega} - \sigma_Q \frac{E}{\rho} \right)-\upr \sigma_Q \frac{B}{\rho} +\upl &=0 \, ,\\
\label{dim_fluid2}   u_t \left( \frac{1}{\omega_B \tau} \upl +\frac{E}{B}+ \upr \tilde{\omega} \right) -\upl \sigma_Q \left(  \frac{E^2}{\rho B}-\frac{B}{\rho} \right) + \upr &= 0 \, .
\end{align}
At this point we can see that the system does not exhibit time reversal, $\omega\rightarrow-\omega$ symmetry. This is to be expected since the magnetic field implicitly breaks time reversal symmetry. 

\subsection{Phase Diagram}

Let us consider global perturbations to the fluid velocities as a probe of stability.  To do so, consider perturbations of the form

\begin{align}
\delta u_x&= (\delta_\perp \sin(2 \omega t)+\delta_\parallel \cos(2 \omega t) + \delta_x)e^{-i w t} \\
\delta u_y &= (-\delta_\perp \cos(2 \omega t)+\delta_\parallel \sin(2 \omega t) + \delta_y)e^{-i w t} \\
\delta \rho &= (\delta \rho_s\sin(\omega t)+ \delta \rho_c \cos(\omega t))  e^{-i w t} 
\end{align}
We can solve the Equations \eqref{T_cons} and \eqref{J_cons} to first order in $\delta$ for the various coefficients and the frequency $w$.  The particular form of the coefficients is not particularly enlightening, however, the imaginary part of $w$ will tell us whether a particular solution for $\upr$ and $\upl$ is stable.

\begin{figure}
\includegraphics[width=3in]{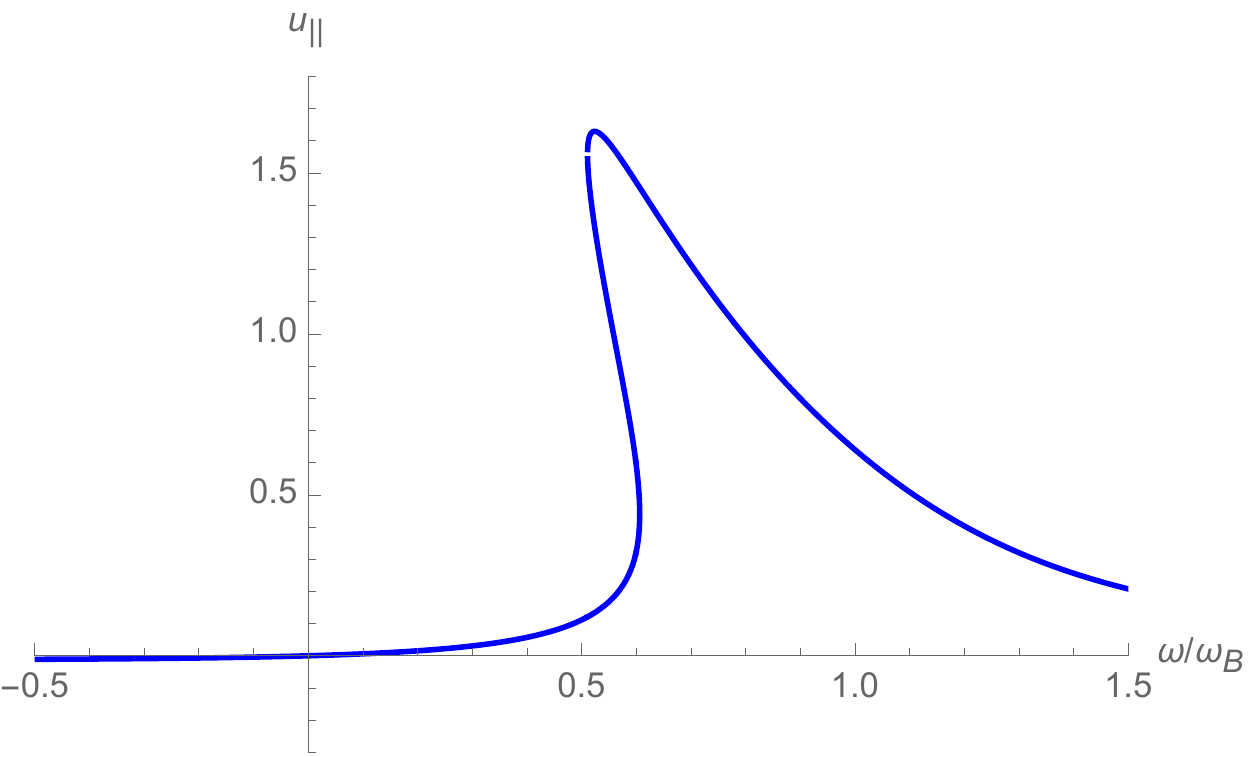}\hfill 
\includegraphics[width=3in]{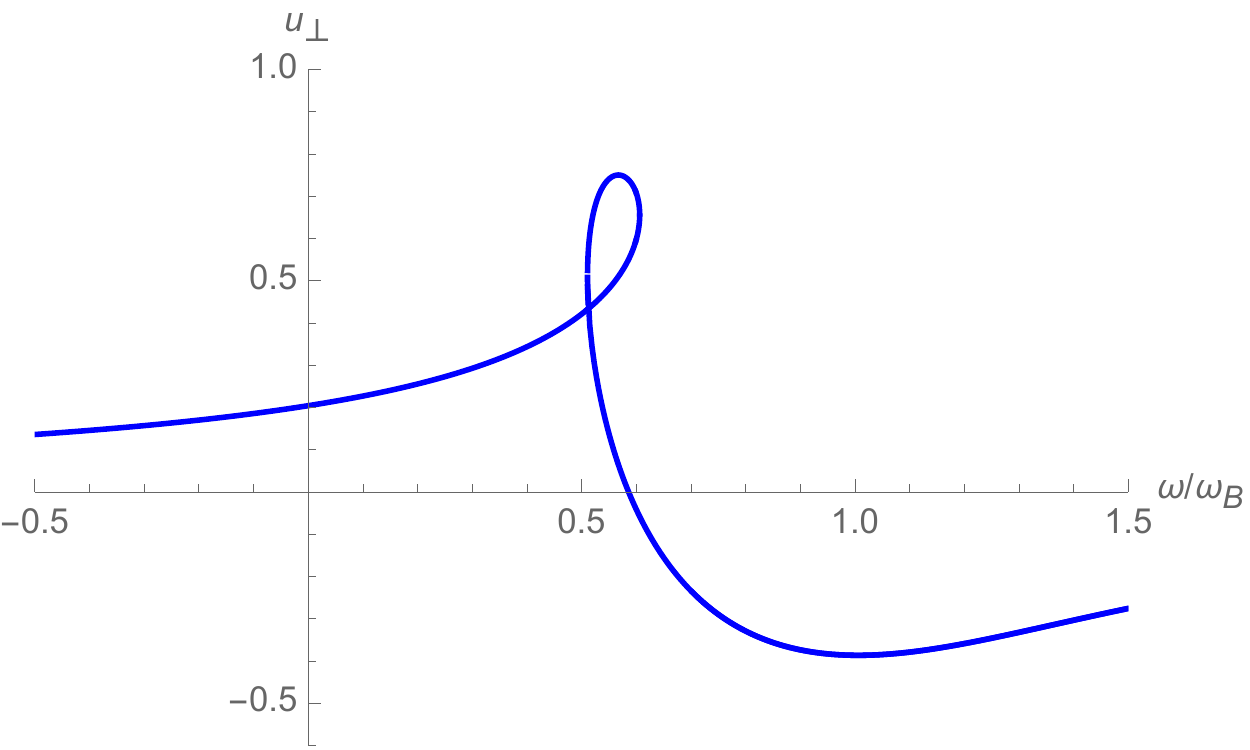}
\caption{
The profile for $\upl$ as a function $\omega$ when $B=1$,  $\omega_B = \rho B/(\epsilon+p)$, $\sigma_Q=0.25$ and $\rho=1$.  We can see that the transition from one solution to three solutions occurs when $\upl'(\omega) = \pm \infty$.
}
\label{fig:upl_trace}
\end{figure}

\begin{figure}
\includegraphics[width=3in]{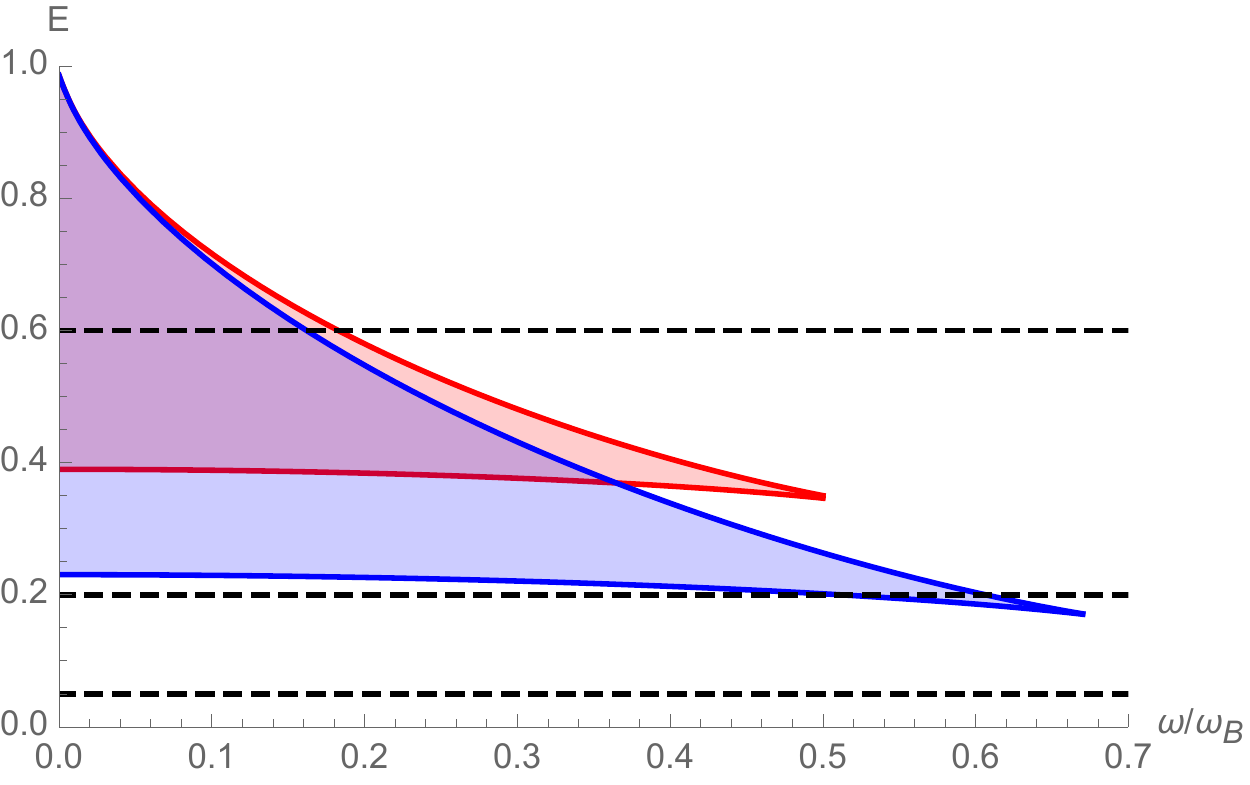}
\hfill
\includegraphics[width=3in]{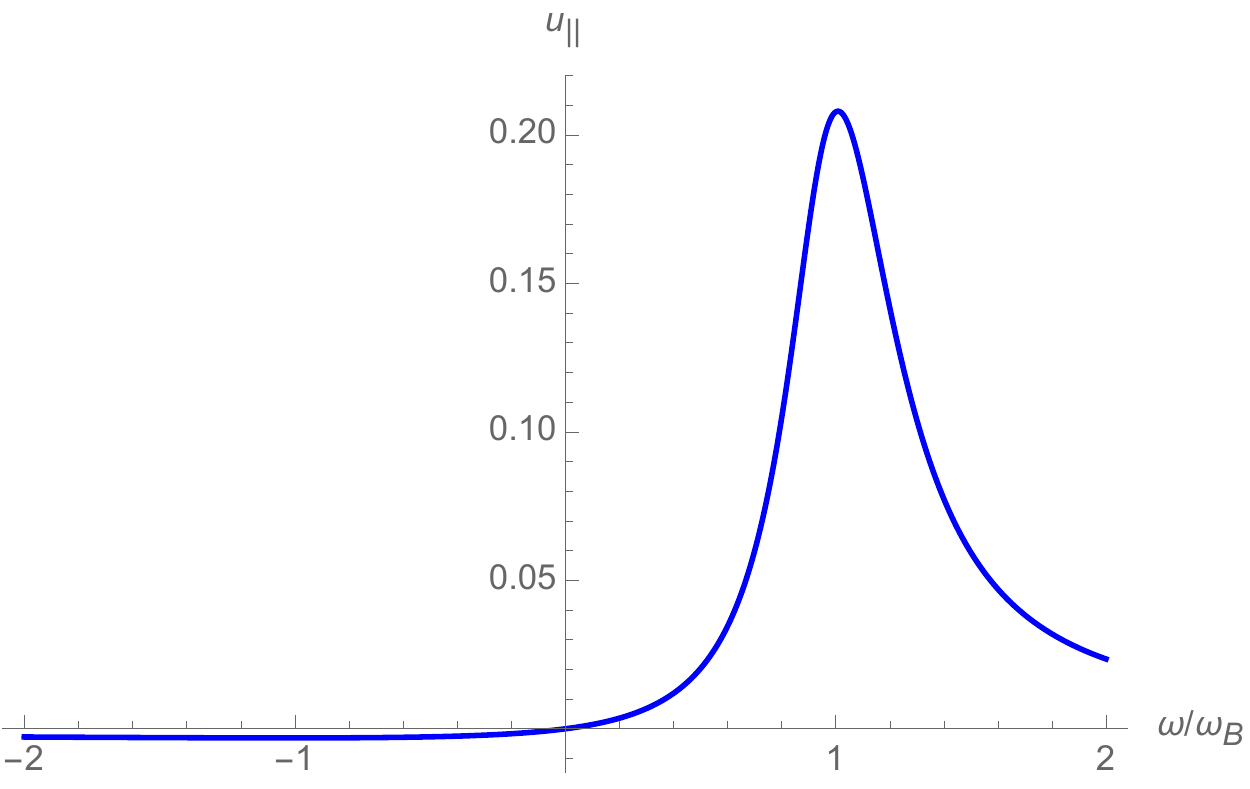}
\\
\includegraphics[width=3in]{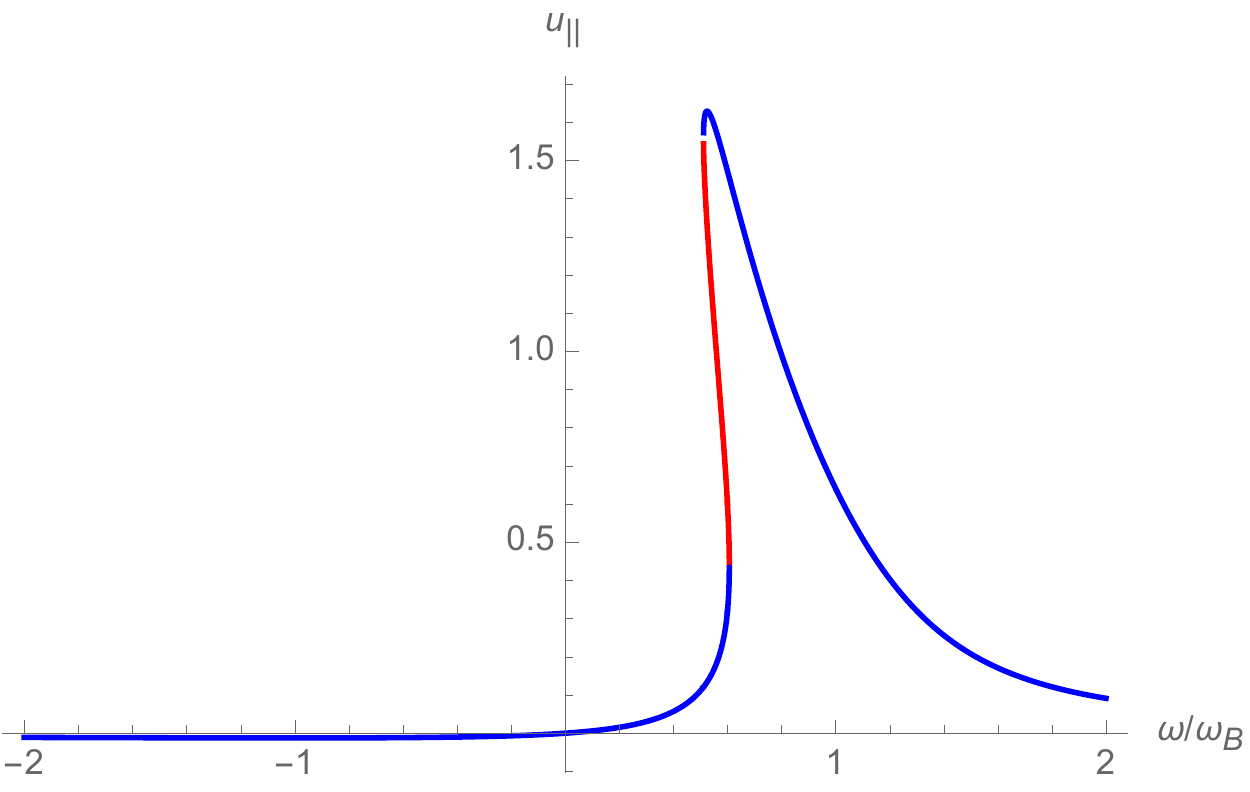}
\hfill
\includegraphics[width=3in]{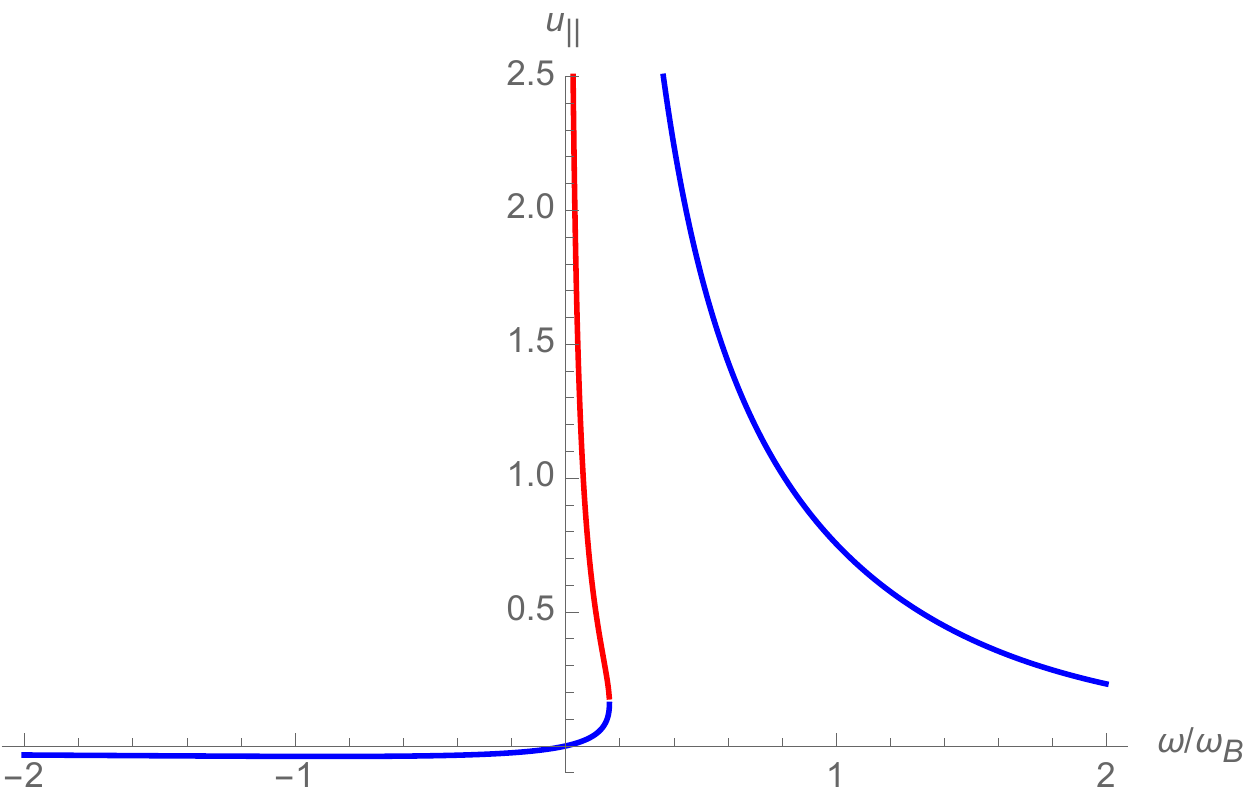}
\caption{
Top left: The phase diagram for $B=1$, $\rho=1$,  $\omega_B = \rho B/(\epsilon+p)$, $\sigma_Q=0.25$ (blue) and $\sigma_Q=0.5$ (red).  In the shaded region there are three solutions of which two are stable; outside there is a single real solution.  The dashed lines corresponds to the value of $E$ in the other plots, ($E=$ 0.05 (top right), 0.2 (bottom left), and 0.6 (bottom right)).  For the bottom two panels the red curve corresponds to the unstable solution.
}
\label{fig:phase}
\end{figure}

Equations \eqref{fluid1} and \eqref{fluid2} are nonlinear with multiple solutions.  We are interested in how many real solutions there are.  For a region in $E$ and $\omega$ plane there are three real solutions, two of which converge and then become imaginary leaving only one real solution in the rest of the $E$ and $\omega$ plane.  In figure \ref{fig:upl_trace} we see that the boundary between one and three solutions is occurs when $\upl'(\omega)=\pm \infty$.  Taking a partial derivative of Equations \eqref{fluid1} and \eqref{fluid2} with respect to $\omega$ and solving for $\upl'(\omega)$ yields

\begin{equation}
\upl'(\omega) = -\frac{u_t \left(u_{\perp} \upl \frac{E}{B}+u_{\perp}^2 \sigma_Q \frac{E}{\rho}+\upl u_{\perp}+u_{\perp} u_t \sigma_Q \frac{B}{\rho}+\upl u_t^2 \tilde{\omega} \right)}
{\omega_B f(\tilde{\omega})} \, ,
\end{equation}
where
\begin{align}
f(\tilde{\omega}) &= u_t \tilde{\omega}^2 + \left(\upl u_{\perp} \sigma_Q \frac{E^2}{\rho B} - \upl u_t \sigma_Q \frac{E}{\rho}  +u_t \upl \frac{E}{B} +1 \right) \tilde{\omega}\\ 
&+u_{\perp}\sigma_Q^2
\frac{E}{\rho} \left( \frac{E^2}{\rho B}-\frac{B}{\rho} \right) +u_t \sigma_Q^2 \left( \frac{B^2}{\rho^2} -\frac{E^2}{\rho^2} \right) +u_\perp \frac{E}{B} +u_t .
\end{align}
So the transition occurs when the $f(\tilde{\omega})=0$.  We can then combine that equation with Equations \eqref{fluid1} and \eqref{fluid2} to determine the boundaries of the region with 3 real solutions.  The ``phase" diagram is presented in figure \ref{fig:phase}.

Using our solution for $w$ we can determine which solutions are stable.  We numerically evaluated the solution along the various portions of the curve. We are not able to prove analytically which solutions are unstable.  However, performing numerical analysis we believe that the middle solution is always unstable.

\begin{figure}
\includegraphics{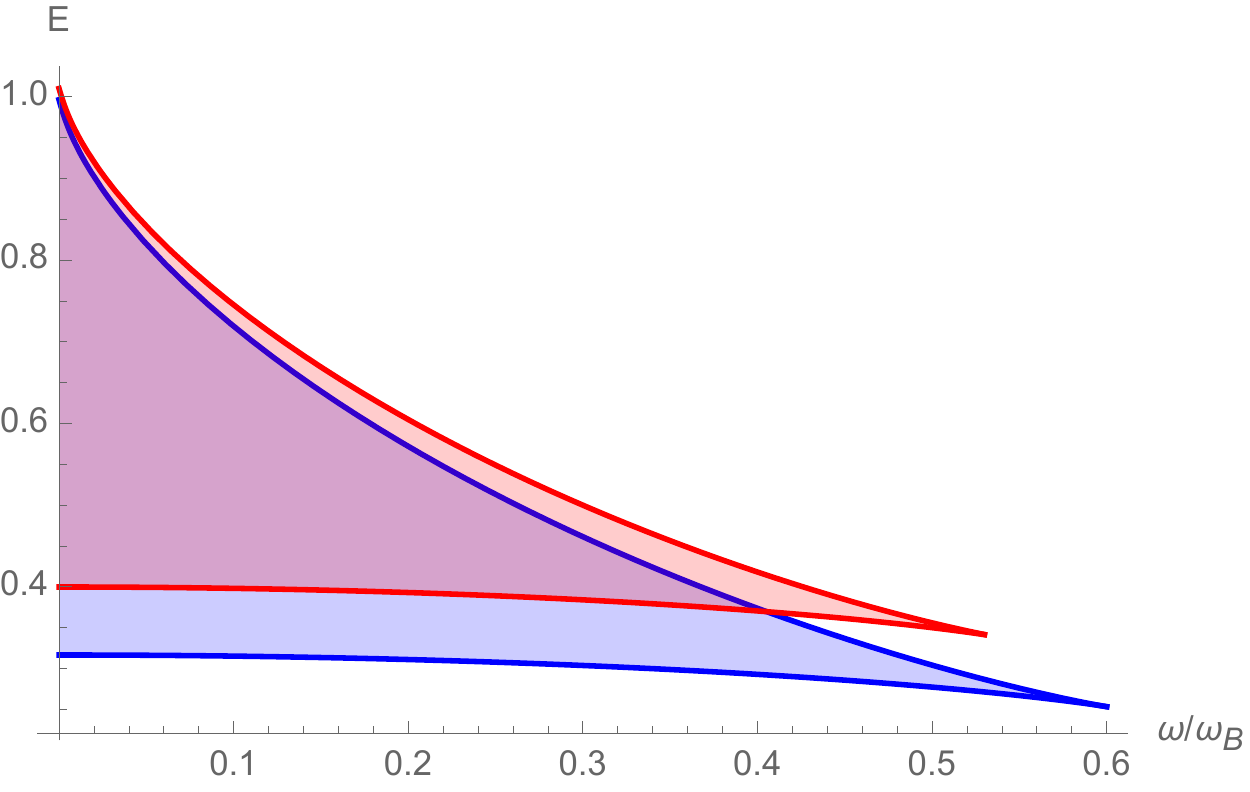}
\caption{
$B=1$, $\sigma_Q=0.25$, $\rho=1$,  $\omega_B = \rho B/(\epsilon+p)$ with $\tau = 5$ (red) $\tau = 10$  (blue).   There is also a region not pictured with 3 real solutions, however, 2 are unstable.  
}
\end{figure}

\subsection{$\sigma_Q =0$}

For the general case we cannot find a simple form for the phase boundary. However, let us consider the simpler case where $\sigma_Q = 0.$  Recall in this case $\upl=0$ and the system of equations simplifies to equation \eqref{sQ0}. As in the general case the phase boundary occurs when $\upr'(\omega)=\pm \infty$. This is given by the solution to 

\begin{align}
\left(2 \upr^2+1\right) \omega (\epsilon+p)+\upr E \rho=\sqrt{1+\upr^2} B \rho \, .
\end{align}

The solution to this and Equations \eqref{sQ0} is 
\begin{align}
\omega = \frac{\rho}{\epsilon+p}\left(B^{2/3}-E^{2/3}\right)^{3/2}\, .
\end{align}

The frequency of the perturbations is given by

\begin{align}
\label{decay} w =\omega -\frac{\sqrt{u_t^2 \tilde{\omega } \left(\xi  \tilde{\omega }+2 B
   \rho u_t\right)-B \rho^2 \left(B \left(\upr^2 (v_s^2
   +1)+1\right)-\upr E (v_s^2 +1) u_t\right)}}{\sqrt{\xi } u_t
   (\epsilon+p)} \, ,
\end{align}
where $\xi = \upr^2(v_s^2-1)-1$, $\tilde{\omega}=(\epsilon+p)\omega$ and $v_s =\sqrt{\partial p/\partial\epsilon}$ is the velocity of sound.  The solutions to equation \eqref{sQ0} are a subset of the solutions to a 4th order polynomial. Plugging in the solutions from that polynomial into \eqref{decay} allows us to check that for $\sigma_Q =0$ the middle solution is indeed the unstable one as the numerical evidence suggests for the general case.

\section{Time Averaged DC conductivity}
Given the stability analysis above, we are now in a position to extract the time-averaged DC conductivities as a function of $\omega$\footnote{Time averaging allows us to identify features of DC transport seen over a large number of periods as would be done in a lab.}. To do so, we subject the fluid to an electric field of the form $\delta E dt\wedge dx$, identify stable solutions, and use the techniques of linear response to extract the conductivity. These usually involve Laplace transforming the linearized hydrodynamic equations and inverting the resulting matrix to find the response functions \citep{HKMS, Kovtun:2012rj}. This method works well for time independent background solutions, where all interesting time dependence is carried by the perturbations. For time dependent background solutions, however, this procedure is no longer tractable. One is left with products of background solutions with unknown functions which can not be decoupled--they turn into convolutions upon Laplace transforming. 

We must then proceed in a brute force manner by solving the perturbative equations. To do so, recall that subjecting the fluid to a magnetic field and an electric field of the form 
\begin{equation}
\vec{E}
=\begin{pmatrix}
E\cos(\omega t) \\
E \sin(\omega t)
\end{pmatrix}
\end{equation}
we find the velocity given by 
\begin{equation} \label{velocity}
\vec{u}=\begin{pmatrix}
 u_{\parallel} & -u_{\perp} \\
 u_{\perp} & u_{\parallel}
 \end{pmatrix}
 \begin{pmatrix} 
 \cos(\omega t) \\
 \sin(\omega t)
 \end{pmatrix}
 = \Lambda \vec{E}
\end{equation}
where $u_{\parallel, \perp}$ are found by solving equations \eqref{fluid1} and \eqref{fluid2} and
\begin{equation}
|| \Lambda_{ik} || = \frac{1}{E}\begin{pmatrix}
 u_{\parallel} & -u_{\perp} \\
 u_{\perp} & u_{\parallel}
 \end{pmatrix}.
\end{equation}
We then plug this solution into the constitutive relations and solve for the internal conductivity of the Floquet fluid:
\begin{equation}
\Sigma_{ij} = \rho\Lambda_{ik}+\sigma_{Q}\delta_{ik}+B\sigma_{Q}\e^{ij}\Lambda_{jk}.
\end{equation}

Now, let us perturb this system via a constant electric field in the $x$ direction: $F\to F+\delta E dt\wedge dx$. To be consistent with the framework of hydrodynamics, we must perturb the hydrodynamic variables. Assuming the background temperature is constant, we have 
\begin{align*}
\mu(r,t) &= \mu+\delta\mu(r,t) \\
\epsilon(r,t) &=\epsilon+\delta \epsilon  \\
u_{i} & =\Lambda_{ij}E_{j}+\delta v_{i}
\end{align*}
as well as 
\begin{align*}
\rho(r,t)&=\rho+\delta \rho \equiv\rho+\frac{\partial \rho}{\partial\mu}_{| T}\delta \mu+\frac{\partial \rho}{\partial T}_{|\mu} \delta \epsilon\\
T(r,t) &=T+\delta T(r,t) \equiv T+\frac{\partial T}{\partial\mu}_{|T}\delta\mu+\frac{\partial T}{\partial \epsilon}_{|\mu}\delta \epsilon \\
P(r,t) &=P+\delta P\equiv P+\rho\delta\mu-c_s^2\delta \epsilon .\\
\end{align*}
Perturbations of the stress tensor and current are given by the constitutive relations. 

Assuming everything is independent of spatial coordinates, the linearized hydrodynamic equations of motion are
\begin{align}
\del_{t}\delta \rho &=- \frac{\sigma_{Q}}{u_0} \del_{t} \left(E^{i}\delta v_{i}+\delta E_{i}\Lambda_{ij}E_{j}\right) \\
\del_{t}\delta \e &= \frac{1}{u_{0}^{2}} \left( \rho E_{i}\delta v_{i}+\delta E_{i}\Sigma_{ij}E_{j}\right) \\
\del_{t}\delta T^{ti} & = -\left(E_{i}\delta\rho+\delta E_{i} \rho\right)+\frac{1}{u_{0}}B\e^{ij}(\rho\delta v_{j}+\delta \rho\Lambda_{jk}E_{k}).
\end{align}

Motivated by the previous section, we choose the following ansatz
\begin{align}
\label{delta1}\delta u_x/\delta E &= \delta_\perp \sin(2 \omega t)+\delta_\parallel \cos(2 \omega t) + \delta_x \\
\label{delta2}\delta u_y/\delta E &= -\delta_\perp \cos(2 \omega t)+\delta_\parallel \sin(2 \omega t) + \delta_y  \\
\delta \epsilon/\delta E &= e_s \sin(\omega t)+e_c \cos(\omega t)\\
\delta \rho/\delta E &= \rho_s \sin(\omega t)+\rho_c \cos(\omega t).
\end{align}
This particular form can be deduced from boost invariance. If we boost to a co-rotating frame, the probe field $\delta E$ appears as if it is rotating with velocity $-\omega$. The response of the fluid in this frame will look like \eqref{velocity} with $\omega \to -\omega$. Boosting back to the lab frame gives us ansatz \eqref{delta1} and \eqref{delta2}. One can think of the $2\omega$ dependence as stemming from the response of the fluid to both the probe and internal $E$ fields, while the constant shift characterizes the response to the probe alone. 

After plugging in \eqref{delta1} and \eqref{delta2} to the hydrodynamic equations, we take the time average over one period. This time average restores the rotational invariance in the plane so the form of the conductivity tensor is determined by two functions $\sigma_{xx}$  and $\sigma_{xy}$, the longitudinal and Hall conductivities respectively

\begin{figure}

\includegraphics[scale=.85]{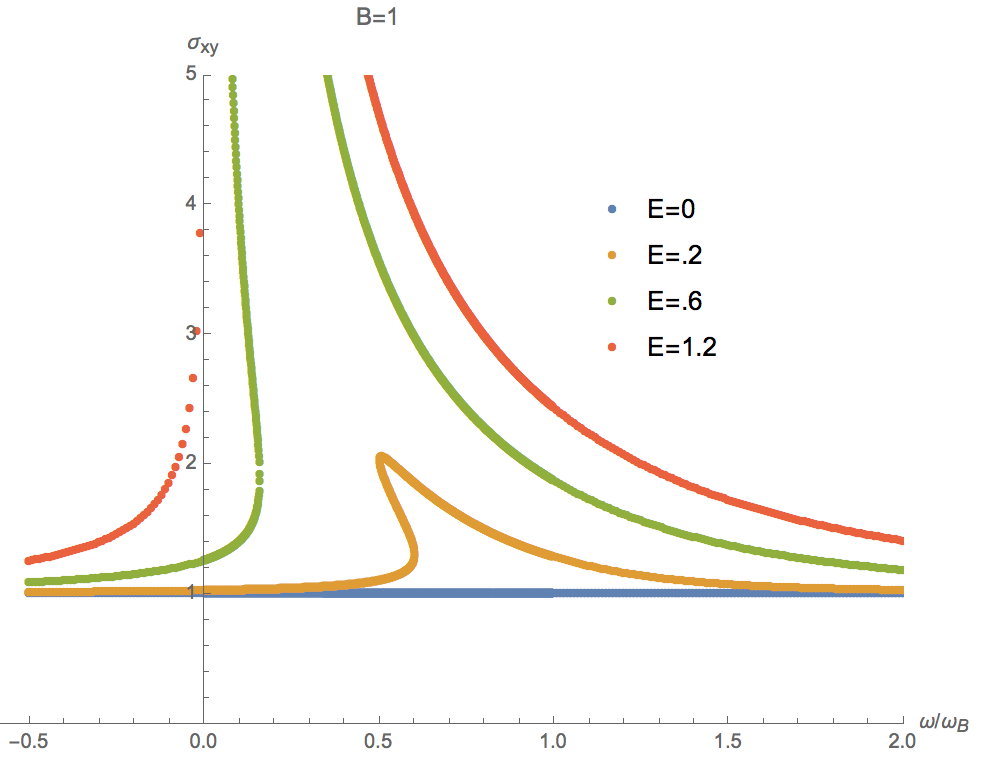}
\caption{
Hall conductivities for various values of $E$ with $B=1$, $\sigma_Q=0.25$, $\rho=1$ and $\epsilon=1$. Each curve asymptotes to $\rho/B$.}
\label{fig:conduct}
\end{figure}

\begin{align}
\bar{\sigma} = \left(
\begin{array}{cc}
 \sigma _{xx} & \sigma _{xy} \\
 -\sigma _{xy} & \sigma _{xx} \\
\end{array}
\right) \, ,
\end{align}
where
\begin{align}
\sigma_{xx}&=\frac{u_t (\upl \rho_c-\upr \rho_s+2 \delta_x \rho)-\sigma_Q
   (\upl E (\delta_\parallel+\delta_x)+\upr E
   (\delta_y-\delta_\perp)+2 u_t (B \delta_y+u_t))}{2u_t}\, ,\\
\sigma_{xy}&=\frac{E \sigma_Q (\upl
   (\delta_\perp+\delta_y)+\upr (\delta_\parallel-\delta_x))-u_t (\upl
   \rho_s+\upr \rho_c+2 B \delta_x \sigma_Q+2 \delta_y \rho)}{2u_t} \, .
\end{align}
The unknown coefficients appearing in \eqref{delta1} and \eqref{delta2} can be found from the inverse of a known $8\times8$ matrix.  This is intractable to write down so we in general solve numerically. This is easily done, since the linearized hydrodynamic equations can be inverted for given values of $E,B,\omega,\epsilon,p,\rho$ and $\sigma_Q$. The specifics of the EOS are not necessary for our work as the fluid is in equilibrium. For our numerical analysis, however, we assume our fluid is conformal and so our EOS is $P=\epsilon/2$.

Results for the Hall conductivities are given in \ref{fig:conduct}. Interestingly, the longitudinal component of the conductivity $\sigma_{xx}$ is zero for all finite values  of $E,B$ and $\omega$, in the DC limit.\footnote{This would not be the case, for example, in the $B \to 0$ limit. However, our numerics fail in this regime since the equations become singular in this limit. A more careful analysis is needed to smoothly take such a limit.} For AC currents $\sigma_{xx}$ is in general non-zero.  

Also visible in figure \ref{fig:conduct} are the same instabilities found in figure \ref{fig:upl_trace}. This is unsurprising, since the conductivity depends on the fluid velocity though the constitutive relations. These instabilities can be interpreted as separating two distinct regions of fluid behavior characterized by monotonically increasing and decreasing conductivities as a function of driving frequency. The width of these regions, as well as the width of the unstable region, is determined by $\sigma_Q$ and the relative magnitude of $E$ and $B$ as can be seen in figure 2. As one increases the driving frequency, the fluid will undergo a phase transition between the phases characterized by a sudden jump in the conductivity. For a fixed value of $B$, the instabilities become increasingly dramatic with increasing $E$, pushing the conductivities (and fluid velocities) to larger and larger values after the phase transition. Similarly, the region of monotonically increasing conductivity shrinks to zero as $E\to B$, until this region is no longer accessible. At that point, the fluid has infinite conductivity for $\omega \to 0$, and monotonically decreasing conductivity for all values of $\omega>0$.

\subsection*{ Behavior at large and small  $\boldsymbol{\omega}$}
Let us now examine the limiting behavior of the Hall conductivities. In the large frequency limit we notice interesting universal behavior.  While at small frequencies there is a market difference between $E<B$ and $E>B$ (see figure \ref{fig:conduct}), at large frequency the two converge. Generically, we have

\begin{align}
\sigma_{xy} \sim \frac{\rho}{B}+\frac{E^2 \sigma_Q^2}{\omega (\epsilon+p)}+\frac{E^2 \rho \left(5 B^2 \sigma_Q^2+\rho^2\right)}{2 B \omega ^2(\epsilon+p)^2}+O(\omega^{-3}) \, .
\end{align}

The leading behavior is the conductivity of a fluid where $E=0$ \citep{HKMS,dyonic}.  This is not surprising given the analysis of section \ref{sec:example}. In that case, we found that in the large frequency limit, a massive relativistic particle's velocity tends to zero. Heuristically this is because the electric field is rotating so fast that the particle can not react fast enough.

We can also consider the small $\omega$ limit for $\sigma_Q=0$, the linear term for $\sigma_Q\neq 0$ was intractable.  Recall that for $E>B$ at $\omega = 0$ there is no solution.  So for $E<B$
\begin{align}
\sigma_{xy} \sim \frac{\rho}{\sqrt{B^2-E^2}}+\frac{E^2 (\epsilon+p)\omega}{(B^2-E^2)^2}+O(\omega^{2}) \, 
\end{align}
which simply is the Hall conductivity for a fluid in a static electric and magnetic field. 

\section{Conclusion}

In this work, we investigated a periodically driven relativistic fluid.  Under some simplifying assumptions, such as the existence of a heat bath in which to absorb energy and a mechanism by which to dissipate momentum, we were able to reduce the system of nonlinear differential equations to a system of algebraic equations.  Analyzing the results we found interesting behavior as a function of amplitude and frequency similar to previous work on strongly interacting Floquet systems \citep{Rangamani2015}.  We also found a similarity to a simple nonlinear driven harmonic oscillator. Namely, there are regions in the $E$, $\omega$ plane for which multiple stable solutions exist. These solutions are connected by an unstable solution.  We wonder if this is a more generic feature of driven nonlinear systems.

We also considered a DC probe electric field and calculated the response of the fluid.  In doing so we are able to measure the conductivity of the system.  This conductivity should be related to the conductivity of some condensed matter systems, such as the cuprates or the Dirac fluid phase of graphene \citep{GrapheneHydro}. In fact, a more traditional Floquet analysis of graphene has been carried out in \citep{Graphene}. There, the authors found that the oscillating field opens a gap in the spectrum and induces a DC Hall conductivity even in the absence of a magnetic field. However, their analysis did not include the electron-electron Coulomb interactions, which are crucial to the formation of the the Dirac fluid phase, nor did they turn on an external magnetic field. These extra complications might require one to abandon the standard band topology approach to Floquet systems, but recent efforts in classifying \emph{interacting} Floquet systems might offer useful insights \citep{IntFloquet1, IntFloquet2}. Regardless, one can ask if a similar analysis to \citep{Graphene} in the presence of these interactions is tractable, and whether the results presented in this paper can be recovered by studying the appropriate ballistic to hydrodynamic crossover regime as in \citep{BtoH}. We leave this for future work.

\section*{Acknowledgments}
We would like to thank Andreas Karch, Larry Yaffe and Andy Lucas for useful discussions and comments on this manuscript. This work was supported, in part, by the U.S. Department of Energy under Grant No. DE-SC0011637.
\begin{appendix}

\section{Viscosity}

We can actually include viscous terms and still obtain fully non-linear solutions.  It is interesting that while the fluid is under going uniform motion there are still contributions coming from the shear viscosity, although the bulk viscosity does not contribute.  We use the following definitions for the shear viscosity.  The projector is given by

\begin{eqnarray}
\Delta^{\mu\nu} \equiv g^{\mu\nu} + u^\mu u^\nu \ .
\end{eqnarray}

The shear tensor itself is written as
\begin{eqnarray}
\sigma^{\mu\nu} \equiv 2 \nabla^{\langle \mu} u^{\nu \rangle} \ .
\end{eqnarray}
Where the angular brackets $\langle \rangle$ force the tensor to be orthogonal to the velocity and tracelessness:
\begin{eqnarray}
A^{\langle \mu \nu \rangle} \equiv \frac{1}{2} \Delta^{\mu\alpha} \Delta^{\nu \beta}(A_{\alpha \beta} + A_{\beta \alpha}) - \frac{1}{2} \Delta^{\mu\nu} \Delta^{\alpha \beta} A_{\alpha \beta} \ .
\end{eqnarray}

The equations of motion governing the fluid are now
\begin{align}
\upr \left(B^2 \sigma_Q \tau +\epsilon+\eta  \tau  \vec{u}^2 \omega
   ^2\right)-\tau  (\upl B q+\upl u_t \omega 
   (\epsilon+p)-B E \sigma_Q u_t)=0 \, ,\\
   \upl \left(\tau 
   \left(\sigma_Q (B^2-E^2) +\eta  \vec{u}^2 \omega
   ^2\right)+\epsilon\right)+\upr \tau  (B q+u_t \omega 
   (\epsilon+p))+E q \tau  u_t = 0 \, ,
\end{align}

where $\eta$ is the viscosity.

\end{appendix}
\bibliographystyle{JHEP}
\bibliography{floquet}

\end{document}